\documentclass[aps,prl,twocolumn,groupedaddress,amsfonts,showpacs]{revtex4}
\usepackage{graphicx,color}
\usepackage{bm}

\begin{document}

\title{Tunneling Anisotropic Magnetoresistance and Spin-Orbit
Coupling in Fe/GaAs/Au Tunnel Junctions}

\author{J. Moser}
\email{juergen.moser@physik.uni-regensburg.de}
\author{A. Matos-Abiague}
\author{D. Schuh}
\author{W. Wegscheider}
\author{J. Fabian}
\author{D. Weiss}

\affiliation{Institut f\"{u}r Experimentelle und Angewandte Physik,
Universit\"{a}t Regensburg, 93040 Regensburg, Germany}

\date{\today}

\begin{abstract}
We report the observation of tunneling anisotropic magnetoresistance
effect (TAMR) in the epitaxial metal-semiconductor system
Fe/GaAs/Au. The observed two-fold anisotropy of the resistance can
be switched by reversing the bias voltage, suggesting that the
effect originates from the interference of the spin-orbit coupling
at the interfaces. Corresponding model calculations reproduce the
experimental findings very well.
\end{abstract}

\pacs{73.43.Jn, 72.25.Dc, 73.43.Qt}%
\keywords{TAMR, tunneling anisotropic magnetoresistance, Fe/GaAs/Au,
spin-orbit coupling}

\maketitle

Tunneling magnetoresistance (TMR) devices consist of a tunneling
barrier, typically an oxide, sandwiched between two ferromagnetic
layers of different coercive fields. Such systems find widespread
use in sensor and memory application as they exhibit a large
resistance difference for parallel and antiparallel alignment of the
ferromagnets' magnetization \cite{zutic}. The TMR effect relies,
within the simplest model \cite{julliere}, on the different spin
polarizations at the Fermi energy E$_{F}$ in the ferromagnets; it is
absent if one ferromagnetic layer is replaced by a normal metal.
Hence it came as a surprise that a spin-valve-like tunnel
magnetoresistance was found in (Ga,Mn)As/alumina/Au sandwiches
\cite{gould}. The origin of the effect, labeled tunneling
anisotropic magnetoresistance (TAMR), was associated with the
anisotropic density of states in the ferromagnet (Ga,Mn)As. An
enhanced anisotropic magnetoresistance (AMR) effect measured across
a constriction in a (Ga,Mn)As film was ascribed to the TAMR effect,
too \cite{giddings}. In both experiments the fourfold symmetry,
expected if the (Ga,Mn)As hole density of states is involved, was
broken and ascribed to strain in (Ga,Mn)As.

Here we show that a TAMR effect can also be observed in sandwiches
involving a conventional ferromagnet like iron. A stack of Fe, GaAs
and Au, with iron grown epitaxially on the GaAs tunneling barrier,
shows pronounced spin-valve-like signatures. We observe a uniaxial
anisotropy of the tunneling magnetoresistance. Depending on the bias
voltage the high resistance state is either observed for the
magnetization \textbf{M} oriented in [110] or in [$\bar{1}$10]
direction. We propose a theoretical model in which the $C_{2v}$
symmetry, resulting from the interference of Bychkov-Rashba and
Dresselhaus spin-orbit interactions, is transferred to the tunneling
probability, giving rise to the observed two-fold symmetry.

\begin{figure}
\includegraphics[width=8cm]{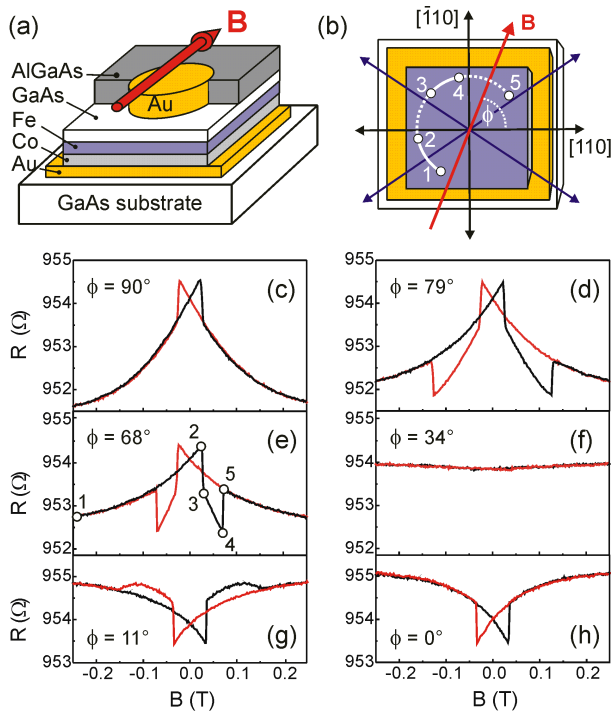}
\caption{\label{fig1}(a) Schematic picture of a Fe/GaAs/Au tunnel
structure; (b) schematic top view on the iron layer; (c)-(h)
tunneling resistance depending on the magnetic field \textbf{B}
swept under different angles $\phi$ measured at 4.2 K and -90 mV
bias (Au contact grounded). The double step switching mechanism is
illustrated in (b) and (e) (see text).}
\end{figure}

A sketch of the system is shown in Fig.~\ref{fig1}(a). The 13 nm
thick epitaxial iron layer was grown on an 8 nm thin GaAs (001)
barrier by transferring the freshly grown GaAs heterojunction from
the molecular beam epitaxy chamber to a magnetron sputtering system
without breaking the ultrahigh vacuum (UHV). The quality of the
interface of a sample from the same wafer was checked by
high-resolution transmission electron microscopy \cite{moser}. The
Fe layer was covered by 50 nm cobalt and 100 nm gold which serves as
back contact. The wafer then was glued upside down to another
substrate and the original substrate was etched away. Finally, the
circular, 150 nm thick top gold contact was made by employing
optical lithography, selective etching of AlGaAs, and UHV magnetron
sputtering. At the Fe/GaAs and the Au/GaAs interfaces Schottky
barriers form. The barrier heights can depend on the preparation
technics \cite{wang1} and were assumed to be 0.75 eV on each side,
which was found for the Fe/GaAs interface \cite{kreuzer}. Hence the
GaAs layer constitutes a nearly rectangular barrier allowing, e.g.,
observation of the TMR \cite{kreuzer, moser}. In total, four batches
of samples which differ in the preparation of the Au layer (with and
without H$^{+}$-plasma etching step, see e.g. Ref. \cite{moser}) or
in an additional annealing step (150$^{\circ}$ Celsius for 1 hour)
were investigated. As the described features are essentially
independent of these details we focus on the results of one sample
(annealed without H$^{+}$-plasma etching) below.

The measurements were carried out at 4.2 K using a variable
temperature insert of a $^{4}$He-cryostat with a superconducting
coil to generate the external magnetic field \textbf{B}. We used a
Semiconductor Analyzer HP 4155A to probe the resistance drop across
the GaAs barrier in four-point configuration. Therefore the top Au
contact was grounded. To vary the direction \textbf{M} of iron, the
sample was mounted in a rotatable sample holder enabling a
360$^{\circ}$ in-plane rotation of \textbf{B}. The direction of
\textbf{B} is given by its angle with respect to the hard [110]
direction (nomenclature with respect to GaAs crystallographic
directions). The I-V-characteristics, measured between top gold and
bottom Fe contact, is strongly nonlinear (not shown). This suggests
that electron transport through the barrier is, as in previous TMR
experiments, dominated by quantum mechanical tunneling
\cite{kreuzer}.

Our Letter is about the anisotropy of the tunneling resistance with
respect to the in-plane magnetization \textbf{M} of the iron
contact. Epitaxial iron has both a cubic anisotropy of bulk iron as
well as an uniaxial contribution stemming from the interface.
Magnetization reversal for an in-plane magnetic field typically
takes place in two steps explained by nucleation and propagation of
90$^{\circ}$ domain walls \cite{cowburn}. Figs.~\ref{fig1}(c) -
\ref{fig1}(h) display the tunneling resistance as a function of
magnetic field \textbf{B} swept in different in-plane directions at
a bias voltage of -90 mV and a temperature of T = 4.2 K.
Fig.~\ref{fig1}(c) shows the resistance for the magnetic field swept
at an angle of $\phi=90^{\circ}$([$\overline{1}$10] direction) from
negative saturation to positive saturation and back. The figure
focuses on the interesting region between -0.25 T and +0.25 T. A
clear spin-valve like signal characterized by one switching event
(one jump in R) is observed for the resistance if \textbf{B} is
applied along this hard direction. If \textbf{B} is applied
11$^{\circ}$ off the hard [$\overline{1}$10] axis the characteristic
second switching process occurs at $\sim$ 0.12 T, as is manifested
in Fig.~\ref{fig1}(d). Decreasing $\phi$ the second switching point
is shifted towards smaller B [Fig.~\ref{fig1}(e)]. This two step
switching process is described in more detail for $\phi=68^{\circ}$
in Fig.~\ref{fig1}(e). Starting close to saturation at -\textbf{B}
[point 1 in Fig.~\ref{fig1}(b)], the average magnetization direction
moves towards the hard magnetic [$\overline{1}$$\overline{1}$0]-axis
(point 2) if \textbf{B} is reversed and increased towards positive
field values. In the first step \textbf{M} switches from near the
easy axis closest to the original direction of \textbf{B} beyond the
easy axis located 90$^{\circ}$ sideways from this one (point 3).
Increasing \textbf{B} further drives \textbf{M} towards the
[$\overline{1}$10] direction (point 4) until, in the second
switching event, \textbf{M} jumps near the easy direction closest to
the new \textbf{B}-direction (point 5). The signal disappears if
\textbf{B} is swept along an easy direction - in the present sample
lying at $\phi=34^{\circ}$  - [Fig.~\ref{fig1}(f)] and changes sign
for \textbf{B} close to [Fig.~\ref{fig1}(g)] or along the hard [110]
direction [Fig.~\ref{fig1}(h)].

Though reminiscent of the AMR effect the results presented here
cannot be explained by the conventional AMR effect of the iron
layer. The resistance change caused by the AMR effect of the iron
layer of only about 4~m$\Omega$ is much smaller than the the
observed change in the tunneling resistance of about 3.5~$\Omega$.
So the AMR effect can be excluded as physical origin of the
measurement and the question for the origin of the anisotropic
resistance remains.

\begin{figure}
\includegraphics[width=8cm]{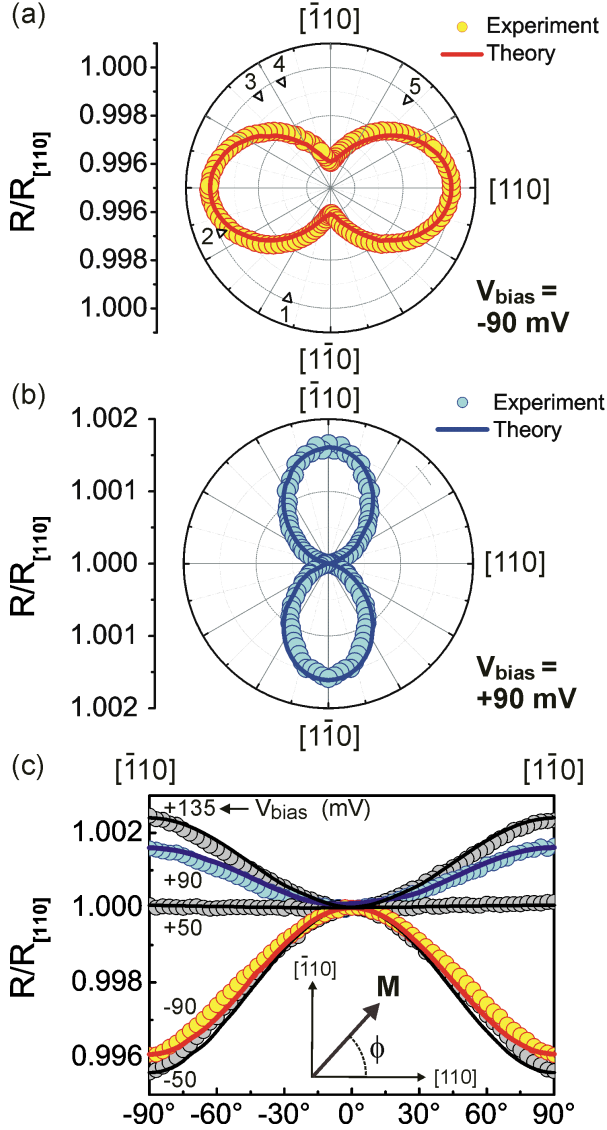}
\caption{\label{fig2}a) $\phi$-scan of the tunneling resistance at
4.2 K and -90 mV bias in a saturation magnetic field
$|$\textbf{B}$|$= 0.5 T and $|$\textbf{B}$|$= 10 T and a theoretical
fit; b) $\phi$-scan at +90 mV; c) $\phi$-scans for different bias
voltages. Symbols correspond to experimental results for -90 mV, -50
mV, 50 mV, 90 mV and 135 mV bias; solid lines correspond to
theoretical results with $\alpha_{l}=42.3~\textrm{eV~\AA}^2,
\alpha_{l}=45.8~\textrm{eV~\AA}^2,
\alpha_{l}=-0.6~\textrm{eV~\AA}^2,
\alpha_{l}=-17.4~\textrm{eV~\AA}^2,
\alpha_{l}=-25.1~\textrm{eV~\AA}^2$ respectively.}
\end{figure}

The symmetry of the anisotropic tunneling magnetoresistance becomes
more explicit at higher \textbf{B} where \textbf{M} is forced to
follow the direction of the externally applied magnetic field. The
data displayed in a polar plot in Fig.~\ref{fig2}(a), normalized to
the resistance in [110] direction, were taken at B = 0.5 T at a bias
voltage of - 90 mV and T = 4.2 K. An uniaxial anisotropy evincing
the shape of a ``horizontal 8" is clearly manifested. The resistance
in [$\overline{1}$10] direction is typically $\sim$ 0.4\% smaller
than in [110] direction. This anisotropy of the resistance explains
the resistance jumps observed in Figs.~\ref{fig1}(c)-\ref{fig1}(h):
The actual position of the (average) magnetization determines the
resistance. The direction highlighted by triangles in
Fig.~\ref{fig2}(a) correspond to the directions, taken up by the
magnetization \textbf{M} in Fig.~\ref{fig1}(e) for the marked
\textbf{B} values. The thin red line is the result of a model
calculation with one adjustable parameter as pointed out below.

The anisotropy depends on the applied bias voltage. If the bias
voltage is reversed from -90 mV to +90 mV the ``8" is rotated by
90$^{\circ}$ as shown in Fig.~\ref{fig2}(b). The bias dependencies
of the resistances' angular characteristics is summarized in
Fig.~\ref{fig2}(c). While for bias voltages V $>$ 50 mV the
resistance is larger for the [$\overline{1}$10] directions, for V
$<$ 50 mV the resistances are largest for the [110] directions.
Similar behavior was found for all samples investigated.

\begin{figure}
\includegraphics[width=8cm]{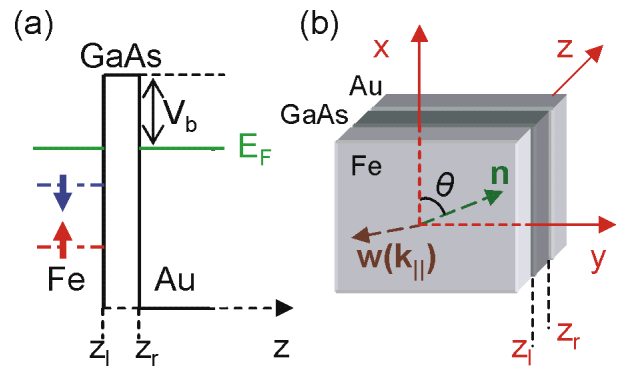}
\caption{\label{fig3}(a) Schematics of the conduction band profile
along the growth direction of the heterostructure. (b) A spacial
view of the model system. The vector $\mathbf{n}$ determines the
magnetization direction in Fe with respect to the $[100]$ direction
($x$ axis). The SOI induced spin precession of the electrons during
tunneling is characterized by the vector
$\mathbf{w}(\mathbf{k}_{\parallel})$ (see text).}
\end{figure}

We now introduce a model ascribing the observed anisotropy to
anisotropic spin orbit interaction. It has already been stated in
reference to GaMnAs junctions
\cite{gould,roster,saito,Shick2006:PRB} that the effect is due to
spin-orbit coupling. However, to capture the $C_{2v}$ symmetry of
the observed TAMR, a uniaxial strain has been invoked \cite{gould}
lowering the four-fold symmetry of the GaMnAs density of states. On
the other hand, ab-initio calculations in CoPt \cite{Shick2006:PRB}
suggest that strain is not necessary to have anisotropic electronic
structure in layered systems. What then leads to the twofold
symmetry of the TAMR? We argue here that TAMR in epitaxial systems
does not need an \textit{ad hoc} anisotropic density of states.
Instead, the tunneling probability itself is strongly anisotropic
due to the interfacial spin-orbit coupling. We propose that the
two-fold symmetry of the TAMR is a consequence of the anisotropic
spin-orbit interaction (SOI) that reflects the bulk and structure
inversion asymmetries of our system. Indeed, the combination of bulk
inversion asymmetry (Dresselhaus SOI)
\cite{dresselhaus,roessler,winkler} and structure inversion
asymmetry (Bychkov-Rashba SOI) \cite{winkler} in GaAs-like
semiconductor heterostructures leads to a spin-orbit interaction
with $C_{2v}$ symmetry. Based on this observation we consider the
following model Hamiltonian for describing the tunnelling across our
metal/semiconductor heterojunction:
\begin{equation}\label{hamilt}
    H=H_{0}+H_{Z}+H_{BR}+H_{D}.
\end{equation}
Here
\begin{equation}\label{h0}
H_{0}=-\frac{\hbar^2}{2}\nabla\left[\frac{1}{m(z)}\nabla\right]+V_{z},
\end{equation}
with $m(z)$ the electron effective mass [in terms of the bare
electron mass $m_{0}$ we assume $m = m_{c} =0.067\; m_{0}$ in the
central (GaAs) region and $m=m_{l}=m_{r} \approx {m_{0}}$ in the
left (Fe) and right (Au) regions] and $V(z)$ the conduction band
profile defining the potential barrier along the growth direction
($z$) of the heterostructure [see Fig.~\ref{fig3}(a)].

The Zeeman spin splitting due to the exchange field (in the Fe
region) and the external magnetic field in the Fe and Au (the Zeeman
energy in GaAs is much smaller than all the other energy scales
characterizing the system and we can therefore neglect its effect)
is given by
\begin{equation}\label{zeeman}
    H_{Z}=-\frac{\Delta(z)}{2}
    \mathbf{n}\cdot \bm{\sigma}.
\end{equation}
Here $\Delta(z)$ represents the Zeeman energy in the different
regions, $\bm {\sigma}$ is a vector whose components are the Pauli
matrices, and $\mathbf{n}$ is a unit vector defining the spin
quantization axis determined by the in-plane magnetization direction
in Fe.

The Bychkov-Rashba SOI due to the structure inversion asymmetry at
the interfaces can be written as \cite{andrada}
\begin{equation}\label{rashba}
H_{BR}=\frac{1}{\hbar}\sum_{i=l,r}\alpha_{i}(\sigma_{x}p_{y}-\sigma_{y}p_{
x})\delta(z-z_{i}),
\end{equation}
where, $\alpha_l$ ($\alpha_r$) denotes the SOI strength at the left
(right) interface $z_{l}=0$ ($z_{r}=d$). We note that inside the
GaAs barrier, away from the interfaces, there is also a
Bychkov-Rashba SOI contribution induced by the applied bias.
However, this contribution is negligible for our system and we
neglect it.

The Dresselhaus SOI resulting from the bulk inversion asymmetry in
GaAs is incorporated in the model through the term \cite{roessler,
winkler,perel,ganichev,wang}
\begin{equation}\label{dresselhaus}
    H_{D}=\frac{1}{\hbar}(\sigma_{x}p_{x}-\sigma_{y}p_{y})
    \frac{\partial}{\partial z}\left(\gamma(z)\frac{\partial}{\partial
    z}\right),
\end{equation}
where the Dresselhaus parameter $\gamma \approx 24 \textrm{ eV
\AA}^3$ in the GaAs region \cite{winkler,perel,ganichev,wang} and
$\gamma = 0$ elsewhere.

The current flowing along the heterojunction is given by
\begin{equation}\label{current}
    I=\frac{e}{(2\pi)^3\hbar}\sum_{\sigma =-1,1}\int
    dE
d^2k_{\parallel}T_{\sigma}(E,\mathbf{k}_{\parallel})[f_{l}(E)-f_{r}(E)],
\end{equation}
where $\mathbf{k}_{\parallel}$ is the in-plane wave vector and
$f_l(E)$ and $f_{r}(E)$ are the electron Fermi-Dirac distributions
with chemical potentials $\mu_{l}$ and $\mu_{r}$ in the left and
right leads, respectively. The particle transmissivity
$T_{\sigma}(E,\mathbf{k}_{\parallel})$ is found, as usually, after
solving for the scattering states in the different regions.

Calculations for the dependence of the resistance on the angle
$\theta$ between the magnetization in Fe and the $[100]$ direction
(note that $\theta =\phi+\pi/4$) were carried out at zero
temperature and a barrier high (measured from the Fermi energy) of
$0.75 \textrm{ eV}$. For the Fe layer we assume a Stoner model with
the majority and minority spin channels having Fermi momenta
$k_{F\uparrow}=1.05 \times 10^8 \textrm{ cm}^{-1}$ and
$k_{F\downarrow}=0.44 \times 10^8 \textrm{ cm}^{-1}$ \cite{jwang},
respectively. The Fermi momentum in Au was assumed to be
$\kappa_{F}=1.2 \times 10^8 \textrm{ cm}^{-1}$ \cite{ashcroft}.

The values of the Bychkov-Rashba parameters $\alpha_{l}$,
$\alpha_{r}$ [see Eq.~(\ref{rashba})] are not known for
metal-semiconductor interfaces. Due to the complexity of the
problem, a theoretical estimation of such parameters requires first
principle calculations including the band structure details of the
involved materials, which is beyond the scope of the present paper.
Here we assume $\alpha_{l}$ and $\alpha_{r}$ as phenomenological
parameters. We have found that due to the large exchange splitting
in the left (Fe) region, the calculated TAMR is dominated by
$\alpha_{l}$; the dependence on $\alpha_{r}$ is negligible and we
can set $\alpha_{r}=0$. This leaves only $\alpha_{l}$ as a fitting
parameter in the comparison of the theoretical and experimental
value of the ratio $R_{[1\bar{1}0]}/R_{[110]}$. Such a comparison is
displayed in Fig.~\ref{fig2}(a) for the case of an external bias
$V_{0}=-90\textrm{ meV}$ and low magnetic field $B=0.5\textrm{ T}$.
The agreement between theory and experiment is indeed very
satisfactory, considering that we fit the value of $\alpha_{l}$ (the
fit is 42.3~$\textrm{eV~\AA}^2)$ only for the direction $\phi =
\pi/2$ --- this is enough for our theoretical model to reproduce the
\textit{complete} angular dependence of $R(\phi)/R_{[110]}$.
Preliminary ab-initio calculations confirm qualitatively the above
picture \cite{popescu}.

We have performed the same fitting procedure for other values of the
applied voltage. The results are shown in Figs.~\ref{fig2}(b) and
(c), where the good agreement between theory and experiment is
apparent. Different values of $\alpha_{l}$ are obtained when varying
the bias voltage, suggesting that the interface Bychkov-Rashba
parameters are voltage dependent (unlike $\gamma$, which is a
material parameter), as found in other systems \cite{zutic}. The
interface Bychkov-Rashba parameter, $\alpha_{l}$, in our system
changes sign at a bias slightly below 50 mV.

The robustness of the fit points to the following phenomenological
model of the TAMR. Averaging the SOI $H_{SOI}=H_{BR}+H_{D}$ [see
Eqs.~(\ref{rashba}) and (\ref{dresselhaus})] over the states of the
system one obtains $H_{SOI} \sim
\mathbf{w}(\mathbf{k}_{\parallel})\cdot \bm{\sigma}$
\cite{ganichev}, where
$\mathbf{w}(\mathbf{k}_{\parallel})=(\tilde{\alpha}
k_{y}-\tilde{\gamma} k_{x},-\tilde{\alpha} k_{x}+\tilde{\gamma}
k_{y},0)$. Here $\tilde{\alpha}$ and $\tilde{\gamma}$ are effective
Bychkov-Rashba and Dresselhaus parameters that measures the SOI
induced spin precession of the electrons during tunneling. There are
only two preferential directions defined by $\mathbf{n}$ and
$\mathbf{w}(\mathbf{k}_{\parallel})$ [see Fig.~\ref{fig3}(b)].
Therefore, the anisotropy of a scalar quantity such as the total
transmissivity is obtained as a perturbative series of $\mathbf{n}
\cdot \mathbf{w}(\mathbf{k}_{\parallel})$, since the SOI is much
smaller than the other relevant energy scales in the system.
Averaging over the in-plane momenta to get the full current, the
anisotropy is determined, up to the second order, by $\langle
[\mathbf{n}\cdot \mathbf{w}(\mathbf{k}_{\parallel})]^2 \rangle$ [the
first-order term vanishes, since
$\mathbf{w}(\mathbf{k}_{\parallel})=-\mathbf{w}(-\mathbf{k}_{\parallel})$]
. Thus, the tunneling current anisotropy is proportional to
$\tilde{\alpha}\tilde{\gamma}\sin 2\theta$ \cite{matos}. Taking into
account that $\theta = \phi + \pi/4$ and that the observed
anisotropy is small one obtains for the TAMR, $R(\phi)/R_{[110]}-1
\sim \tilde{\alpha}\tilde{\gamma}(\cos2\phi-1)$. This is precisely
the kind of angular dependence experimentally found (see
Fig.~\ref{fig2}). Assuming that the spin-orbit parameters are
voltage dependent, one can change the sign and magnitude of the
anisotropy, $\tilde{\alpha}\tilde{\gamma}$, by varying the bias
voltage, as shown in Fig.~\ref{fig2}. Notably, if
$\tilde{\alpha}\tilde{\gamma} \approx 0$, one obtains a suppression
of the TAMR effect, a situation corresponding to a bias voltage of
50 mV [Fig.~\ref{fig2}(c)].

In summery we observed TAMR in an epitaxial ferromagnet/insulator
system and propose that the effect occurs whenever both Rashba and
Dresselhaus SOI are involved in the tunneling process.

Financial support by German Science Foundation (DFG) via SFB 689 and
by BMBF (nanoQUIT) is gratefully acknowledged.



\end{document}